\documentclass[11pt]{article}
\usepackage{cite}
\usepackage{amssymb}
\usepackage{amsmath}
\usepackage{bm}
\usepackage{graphicx}
\usepackage{commath}
\usepackage{esint}
\usepackage{xcolor}	
\usepackage[colorlinks]{hyperref}

\begin{document}

\title{The Dirac-Bohm Picture}
\author{B. J. Hiley\footnote{E-mail address b.hiley@bbk.ac.uk.} and G. Dennis.}
\date{TPRU, Birkbeck, University of London, Malet Street, \\London WC1E 7HX.\\Physics Department, University College London, Gower Street, London WC1E 6BT.}
\maketitle

\begin{abstract}
We examine Dirac's early algebraic approach which introduces the {\em standard} ket and show that it emerges more clearly from a unitary transformation of the operators based on the action.  This establishes a new picture that is unitarily equivalent to both the Schr\"{o}dinger and Heisenberg pictures.  We will call this the Dirac-Bohm picture for the reasons we discuss in the paper.  This picture forms the basis of the Feynman path theory and allows us to show that the so-called `Bohm trajectories' are averages of an ensemble of Feynman paths.

\end{abstract}

\section{Representations and Pictures}
The Stone-von Neumann theorem~\cite{jn31, jn32, ms30} proves that the Schr\"{o}dinger 
representation is unique up to a unitary transformation.  This means that  there could be many equivalent representations or `pictures' as they are often called in this context.  For example the Schr\"{o}dinger picture, where all operators are independent of time, while the wave function or ket carries the time dependence, is well known.  In this picture the Hamiltonian is written in the form
\begin{eqnarray*}
 H_S=H(\hat q,\hat p)\hspace{0.5cm} \mbox{with the wave function }\hspace{0.1cm} \psi(q,t).
\end{eqnarray*}
Here the momentum operator is written in the form $\hat p=-i\hbar\partial/\partial q$.

When considering quantum field theory, it is the Heisenberg picture that comes to the fore.  In this picture all the time dependence is taken into the operators by introducing a unitary transform $U(t)$ so that 
\begin{eqnarray*}
H_H(t)=U^\dag H_SU\hspace{0.3cm}\mbox{with}\hspace{0.3cm} U(t)=e^{iHt/\hbar}
\end{eqnarray*}
where $H(q,p)$ is the Hamiltonian. Then we have the following relations 
\begin{eqnarray*}
\hat q_H(t)=U^\dag \hat q_SU\hspace{0.3cm} \mbox{and}\hspace{0.3cm} \hat p_H(t)= U^\dag \hat p_S U.
\end{eqnarray*}

Apart from the interaction picture and the Fock picture, a little-known  picture was implicitly introduced by Dirac~\cite{pd47}   in which
\begin{eqnarray}
H_D(\hat q_D, \hat p_D)=V^\dag H_S(\hat q_S, \hat p_S)V\hspace{0.3cm}\hspace{0.3cm}\mbox{with}\hspace{0.3cm}V(t)=e^{iS(q,t)/\hbar}	\label{eq:SUTrans}
\end{eqnarray}
where $S(q,t)$ is the classical action. In this case we can then write
\begin{eqnarray*}
\hat q_D=V^\dag \hat q_S V \hspace{0.5cm}\Rightarrow\hspace{0.5cm} \hat q_D=\hat q_S,\hspace{1.5cm}\\
\hat p_D=V^\dag \hat p_S V\hspace{0.5cm}\Rightarrow \hspace{0.5cm}\hat p_D=\hat p_S+\left(\frac{\partial S}{\partial q}\right).
\end{eqnarray*}
To determine the wave function evolution, we use the Schr\"{o}dinger equation written in the form
\begin{eqnarray*}
i\hbar\frac{\partial \psi(q,t)}{\partial t}=H_D(\hat q_D, \hat p_D)\psi(q,t).
\end{eqnarray*}
Utilising $\hat p_D=\hat p_S+\left(\frac{\partial S}{\partial q}\right)$, the quadratic term in the Hamiltonian becomes
\begin{eqnarray*}
\hat p^2_D=\left[\hat p_S+\left(\frac{\partial S}{\partial q}\right)\right]^2=\hat p_S^2+\hat p_S\left(\frac{\partial S}{\partial q}\right)+\left(\frac{\partial S}{\partial q}\right)\hat p_S+\left(\frac{\partial S}{\partial q}\right)^2.
\end{eqnarray*}
If we use $\hat p_S=-i\hbar\partial/\partial q$, we see that $\hat p_D^2$ has real and imaginary parts.  The real part can be written as
\begin{eqnarray*}
\Re(\hat p_D^2)=\hat p_S^2+\left(\frac{\partial S}{\partial q}\right)^2,
\end{eqnarray*}
while the imaginary part becomes
\begin{eqnarray*}
\Im(\hat p_D^2)=-i\hbar\frac{\partial^2S}{\partial q^2}+2\left(\frac{\partial S}{\partial q}\right)\hat p_S.
\end{eqnarray*}
The Schr\"{o}dinger equation can now be expressed in the form
\begin{eqnarray}
i\hbar R(q,t)^{-1}\frac{\partial R(q,t)}{\partial t}-\hbar\frac{\partial S(q,t)}{\partial t}=R^{-1}(q,t)H_D(\hat q_D, \hat p_D)R(q,t)	\label{eq:RSS}
\end{eqnarray}
where we have written the wave function in its polar form 
\begin{eqnarray}
\psi(q,t)=R(q,t)\exp[iS(q,t)/\hbar].	\label{eq:polarpsi}
\end{eqnarray}
In order to see exactly what is going on, choose a Hamiltonian $H=\hat p^2/2m+V(q)$ for simplicity.  The Schr\"{o}dinger equation can then be split into its real and imaginary parts, the real part being
\begin{eqnarray}
\frac{\partial S(q,t)}{\partial t}+\frac{1}{2m}\left(\frac{\partial S(q,t)}{\partial q}\right)^2-\frac{\hbar^2}{2mR(q,t)}\left(\frac{\partial^2R(q,t)}{\partial q^2}\right)+V(q)=0
\label{eq:QHJ}
\end{eqnarray}
which is identical to the quantum Hamilton-Jacobi equation of the Bohm approach~\cite{db52}.

The imaginary part of equation (\ref{eq:RSS}) becomes
\begin{eqnarray}
i\hbar \frac{\partial R(q,t)}{\partial t}=\left[-i\hbar\left(\frac{\partial^2S(q,t)}{\partial q^2}\right)+2\left(\frac{\partial S(q,t)}{\partial q}\right)\hat p_S\right]R(q,t)
\end{eqnarray}
which can be written in the form
\begin{eqnarray}
\frac{\partial\rho(q,t)}{\partial t}+\frac{1}{m}\frac{\partial}{\partial q}\left(\rho(q,t)\frac{\partial S(q,t)}{\partial q}\right)=0.	\label{eq:ConP}
\end{eqnarray}
Here $\rho(q,t)=R^2(q,t)$ is the probability density and the equation is an expression of the conservation of probability.  Thus both the Dirac~\cite{pd47}  approach and the Bohm approach~\cite{db52} lead to the same equations.  For this reason we will call the representation the Dirac-Bohm picture\footnote{We call it the `Dirac-Bohm picture' because, as Schweber~\cite{ss64} points out, the `interaction picture' is sometimes called the `Dirac picture'.}.

\section{Development of the Dirac-Bohm Picture}

In order to derive equations (\ref{eq:QHJ}) and (\ref{eq:ConP}), Dirac~\cite{pd47} used  a new notation 
which, he claims, ``provides a neat and concise way of writing, in a single scheme, both the abstract [sic, non-commuting] quantities themselves and their coordinates, and thus leads to a unification" of the wave picture and the non-commutative algebra of operators~\cite{pd39}. It is this structure that we call the `algebraic approach'.

A key element in this approach was the introduction of a new symbol into the algebra, namely the {\em standard } ket, $\rangle$ [NB without the vertical line~$|$~]\footnote{The symbol $\rangle$ first appears in Dirac~\cite{pd39} and was introduced in his classic text~\cite{pd47}. In later publication \cite{pd12} the symbol appears as $|S\rangle$.  In Frescura and Hiley~\cite{ffbh84} it appears as $|\:\:\rangle$.}.   With this new element, the wave function can be written as a function of {\em operators}.   Consequently all the information usually encoded in the wave function may then be abstracted from the algebra itself, viz, $\psi(\hat q, \hat p)\rangle$ without the need for a Hilbert space.   In order to express the wave function in configuration space, we choose the standard ket  to satisfy the relation $\hat p\;\rangle_q=0$.  This is a natural result since $\hat p \psi\rangle_q=-i\hbar \frac{\partial\psi}{\partial \hat q}\rangle_q$ if we anticipate the Schr\"{o}dinger representation.  Then if $\psi=\mbox{constant}$, we have $\hat p\;\rangle_q=0$. If we write a general function of the set  $\{\hat q,\hat p\}$ in {\em normal order}, we obtain 
\begin{eqnarray*}
\psi(\hat q,\hat p, t)\rangle_q=\psi(\hat q, t)\rangle_q\rightarrow \psi (q,t).
\end{eqnarray*}
By introducing a dual symbol, the {\em standard} bra, $_q\langle\;$, we can formally write
\begin{eqnarray*}
_q\langle q|\psi(\hat q,\hat p,t)\rangle_q=\psi(q,t).
\end{eqnarray*}
In this way we arrive at the usual wave function in a configuration space.

To choose the $p$-representation, we introduce a standard ket satisfying $\hat q\rangle_p=0$  since in this representation $\hat q=i\hbar \frac{\partial}{\partial p}$.  In this case we write the operators in {\em anti-normal order} to obtain any element in the $p$-representation.

Thus one can derive the two equations (\ref{eq:ConP}) and (\ref{eq:QHJ}) directly from the Schr\"{o}dinger equation written in the {\em algebraic} form
\begin{eqnarray}
i\hbar\frac{\partial}{\partial t}\left(Re^{iS/\hbar}\right)\rangle_q=H(\hat q, \hat p)Re^{iS/\hbar}\rangle_q.	\label{eq:ASeqn}
\end{eqnarray}
Dirac  remarked that equation (\ref{eq:ConP}) was similar to the continuity equation used in fluid dynamics but did not mention the earlier work of Madelung~\cite{em26}, who introduced the hydrodynamical model.  
He further noted that equation (\ref{eq:QHJ}) had 
the form of the classical Hamilton-Jacobi equation if one neglected terms of order $\hbar^2$ and above.  Dirac did not pursue this line of reasoning because he thought it would come into conflict with the uncertainty principle.
Full details of this approach can be found in Sections 31 and 32 of his book ``The Principles of Quantum Mechanics", together with comments as to why he did not continue the investigation\cite{pd47}.

\subsection{The Bohm Approach}

Initially Bohm wrote ``Quantum Theory"~\cite{db51} in order to provide an explanation of the standard approach to quantum mechanics--a book in which he argued that hidden variables could {\em not} be used to provide a deeper understanding of the formalism.  However Bohm told one of us [BJH] that after completing the book he felt dissatisfied with his account. 

 While investigating the first order WKB approximation, Bohm noticed, just as Dirac had done, that classical ideas still held.  Unlike Dirac, Bohm saw no reason for abandoning the conceptual structure when higher order terms were retained.  Furthermore Bohm found that he could derive the two equations (\ref{eq:QHJ}) and (\ref{eq:ConP})  in a much simpler way by taking the real and imaginary parts of the Schr\"{o}dinger equation under polar decomposition of the wave function, a process that automatically retains all higher powers of $\hbar$. He went on to show that, contrary to Dirac's conclusion, the uncertainty principle was not violated when retaining higher powers of $\hbar$.

 Bohm's interpretation was based on the assumption that a particle had both a position and a momentum at all times.  He further assumed that the momentum was specified by retaining the canonical relation $p(q,t)=\partial S(q,t)/\partial q$, when the classical action $S(q,t)$ was the phase of the wave function, an assumption that was not justified in Bohm's original papers~\cite{db52}. 
With that assumption Bohm showed that the uncertainty principle would not be violated.  The approach provides a consistent interpretation of a Schr\"{o}dinger particle, providing a clear, intuitive account of all quantum phenomena by removing many of the paradoxes of the standard interpretation~\cite{ph95}. Although it appears to be a return to classical determinism, a closer examination shows
it contains properties that are radically different from classical physics as detailed in Bohm and Hiley~\cite{dbbh93}.

\section{ Quantum Particle Trajectories?}

Perhaps the most contentious aspect of the Bohm approach was the unambiguous appearance of what was regarded as `particle trajectories'.  This follows from the assumption that a particle has a well-defined  position and momentum even though it is not possible to measure them simultaneously. 

 From the fact that position and momentum cannot be measured simultaneously, one can draw two opposite conclusions. Namely,  quantum particles do not have a simultaneous position and momentum, or if they have, we cannot say anything meaningful about them together.  This is the position taken by the majority of physicists.

The opposite assumption, that quantum particles do have a well-defined position and momentum, may have experimental consequences that can be fully explored as they arise.  Hopefully one of the consequences will lead to new experiments.  This is the position adopted by Bohm in his 1952 paper~\cite{db52}.

The fact that one could obtain smooth `trajectories' from a basically non-commutative structure was always a concern. Nevertheless  the appearance of the paper by Kocsis {\em et al.}~\cite{skbbsr11} changed all that.  
This paper provided a method for measuring and constructing such `trajectories'.   These pioneering experiments used single photons but the claim that the momentum flow lines were `photon trajectories' was controversial.  

The further claim that their results vindicated the Bohm approach because the flow lines bore a remarkable similarity to those calculated by Philippidis, Dewdney and Hiley~\cite{cpcd79} unfortunately cannot be sustained~\cite{dmlr15}. 
Photons are zero rest mass particles travelling at the speed of light and therefore cannot
be described by the Schr\"{o}dinger equation.  Flack and Hiley~\cite{rfbh16} have shown that the Kocsis {\em et al.} experiments \cite{skbbsr11} constructed momentum  flow lines,  rather than photon trajectories, and that these flow lines are determined by the weak Poynting vector.

Nevertheless the same techniques can be used on non-relativistic atoms~\cite{bh11,rfbh18} and such experiments are under construction~\cite{jmpe16}.  An analysis of the method shows
 that the experimentally determined curves are not the trajectories of a single atom, but are actually momentum flow lines averaged over many individual single atom events, which is in agreement with the results of Kocsis {\em et al.}~\cite{skbbsr11}.
These new factors call into question the validity of the original assumption that particles follow well-defined trajectories and demand a radical re-evaluation of what underlies the Bohm approach.  Do we return to the common assumption that it is not possible to talk about trajectories of quantum particles~\cite{llel13} or can we probe more deeply? 

   \section{Dirac Trajectories}
   
   What seems to have been forgotten is that Dirac~\cite{pd45} attempted to construct what he called `quantum trajectories' within the context of the non-commutative algebra of quantum operators.  We will bring this out using the Dirac-Bohm picture which avoids the need to introduce the standard ket symbol $\rangle$ and work directly in Hilbert space.  
   
  Following Dirac, we note the general ket evolves in time via
   \begin{eqnarray}
|t\rangle=V(t,t_0)|t_0\rangle.	\label{eq:VT}
   \end{eqnarray}
  We can write
   \begin{eqnarray*}
   \frac{d|t_0\rangle}{dt_0}=\lim_{t\rightarrow t_0}\frac{|t\rangle-|t_0\rangle}{t-t_0}=\lim_{t\rightarrow t_0}\frac{V-1}{t-t_0}|t_0\rangle.
   \end{eqnarray*}
   If we put
   \begin{eqnarray*}
  i\hbar  \lim_{t\rightarrow t_0}\frac{V-1}{t-t_0}= H(t_0),
   \end{eqnarray*}
   we have for general $t$
   \begin{eqnarray*}
   i\hbar \frac{d|t\rangle}{dt}=H(t)|t\rangle.
   \end{eqnarray*}
   Using equation (\ref{eq:VT}) we obtain
   \begin{eqnarray*}
   i\hbar\frac {dV}{dt}|t_0\rangle=H(t)V|t_0\rangle.
   \end{eqnarray*}
   
   Since this holds for any $|t_0\rangle$, we have
   \begin{eqnarray*}
   i\hbar\frac{dV}{dt}=H(t)V.
   \end{eqnarray*}
   Thus for any operator $\hat r$ we have
   \begin{eqnarray*}
   \hat r_t=V^{-1}\hat r V
   \end{eqnarray*}
   so that 
   \begin{eqnarray*}
   i\hbar\frac{d\hat r_t}{dt}=V^{-1}\hat rHV-V^{-1}HV\hat r_{t}=[\hat r_t, H_t]\quad\\\mbox{with}\quad H_t=V^{-1}H V.\hspace{3cm}
   \end{eqnarray*}
   This is a Heisenberg-type equation of motion only now we use a unitary transformation generated by the classical action (\ref{eq:SUTrans}).
   
   Introducing the bra $\langle q_t|=\langle q_{t_0}|V$, we have
   \begin{eqnarray*}
   i\hbar\frac{d}{dt}\langle q_t|=i\hbar \langle q_{t_0}|\frac{dV}{dt}=\langle q_{t_0}|HV=\langle q_t|H_t.
   \end{eqnarray*}
   Thus we can write
   \begin{eqnarray*}
   i\hbar\frac{d}{dt}\langle q_t|q'\rangle=\int \langle q_t|H_t|q''_t\rangle dq''_t\langle q''_t|q'\rangle
   \end{eqnarray*}
   showing that the transition amplitude [TA] $\langle q_t|q'\rangle $ satisfies the Schr\"{o}dinger wave equation.  We can therefore in general write
   \begin{eqnarray*}
   \langle q_t|q'\rangle=e^{iS'(q_t,q';t,t')/\hbar}.
   \end{eqnarray*}
   This then links with equation (\ref{eq:ASeqn}) where we have replaced $Re^{iS/\hbar} $ by $e^{iS'(q_t,q'; t,t')/\hbar}$. Here $S'$ is now a complex number. (For more details see Flack and Hiley~\cite{rfbh18}.)
   
    However for simplicity, let us continue with the special case assuming $S$ real.  In the limit $\hbar\rightarrow 0$, when $S$ is real we have at the initial point $q$
  \begin{eqnarray}
  \frac{\partial S}{\partial t_0}=H_c(q, p)\hspace{0.2cm}\mbox{and}\hspace{0.2cm}p=-\frac{\partial S}{\partial q},	\label{eq:ContA}
  \end{eqnarray}
 where $H_c(q,p)$ is the classical Hamiltonian.   At the final point $q'$ we have
  \begin{eqnarray}
   -\frac{\partial S}{\partial t}=H_c(q', p')\hspace{0.2cm}\mbox{and}\hspace{0.2cm}p'=\frac{\partial S}{\partial q'}.	\label{eq:ContB}
  \end{eqnarray}
  These equations define the connection between the variables $(q,p)$ and the variables $(q', p')$ by a contact transformation.  The unitary transformation given by equation (\ref{eq:lift}) below is the quantum generalisation of the contact transformation, a point that Bohm~\cite{db51} emphasises.
See Brown and Hiley~\cite{mbbh00} for a different discussion of this point.  
    
  \subsection{ The Role of the Transition  Amplitudes} \label{sec:DandF}
  
  In the Dirac-Bohm picture, the operators are functions of time. Strict attention must therefore be paid to the time order of the appearance of elements in a sequence of operators.  In the non-relativistic limit, operators at different times always commute\footnote{In this paper we will, for simplicity, only consider the non-relativistic domain.  Dirac himself shows how the ideas can be extended to the relativistic domain. For a more detailed discussion of the relativistic approach see Schwinger~\cite{js51}.}. 
  Hence  a time-ordered sequence of position operators  can be uniquely written in the  form,
\begin{equation}
	\langle q_t|q_{t_0}\rangle=\int\dots\int\langle q_t|q_{t_j}\rangle dq_j\langle q_{t_j}|q_{t_{j-1}}\rangle\dots \langle q_{t_2}|q_{t_1}\rangle dq_1\langle q_{t_1}|q_{t_0}\rangle.		\label{eq:QTraj}
	\end{equation}
This divides the TA, $\langle q_t|q_{t_0}\rangle$, into a sequence of closely adjacent points, each pair of which is connected by an  infinitesimal TA.  Dirac writes ``\dots one can regard this as a trajectory\dots and thus makes quantum mechanics more closely resemble classical mechanics"~\cite{pd45}.

In order to analyse the sequence (\ref{eq:QTraj}) further, Dirac assumed that for a small time interval $\Delta t=\epsilon$, we can write
\begin{equation}
	\langle q'|q\rangle_\epsilon= \exp[iS_\epsilon(q',q)]		\label{eq:lift}
\end{equation}
where we will again take $S_\epsilon(q',q)$ to be a real function. This, of course, became the source of Feynman's propagators~\cite{rpf48}.  However  Dirac~\cite{pd47} had previously shown that
\begin{equation}
p_\epsilon(q',q)=\frac{\langle q'|\hat P|q\rangle_\epsilon}{\langle q'|q\rangle_\epsilon}=\frac{i\hbar\nabla_{q}\langle q'|q\rangle_\epsilon}{\langle q'|q\rangle_\epsilon}=	-\nabla_{q} S_\epsilon(q',q)		\label{eq:conjM1}
	\end{equation}		
and	
\begin{equation}
 p'_\epsilon(q',q)=\frac{\langle q'|\hat P'|q\rangle_\epsilon}{\langle q'|q\rangle_\epsilon}=\frac{-i\hbar\nabla_{q'}\langle q'|q\rangle_\epsilon}{\langle q'|q\rangle_\epsilon}=	\nabla_{q'} S_\epsilon(q',q).		\label{eq:conjM}
	\end{equation}
Here $\hat P$ is the momentum operator.	These equations replace the classical equations (\ref{eq:ContA}) and (\ref{eq:ContB}) and generate the quantum relation between the points $(q, p)$ and $(q', p')$.  It is this analogy that prompted Dirac to introduce the notion of a particle trajectory.  The details of this relationship are discussed by de Gosson and Hiley~\cite{mdgbh11}.  In that paper it is shown that the quantum trajectories can be formed by lifting the classical trajectories generated by symplectic transformations onto the covering space, the covering group being the metaplectic group and its generalisation.

  \section{Quantum Trajectories and Feynman Propagators}
  
  \subsection {The Momentum Propagator}	\label{sec:MomProp}
  
  Let us continue to discuss the notion of  `quantum trajectories'  based on equation (\ref{eq:QTraj}).   Each TA between two points can be subdivided into a series of infinitesimal TAs $\langle q'| q\rangle_\epsilon$. Furthermore each infinitesimal TA may be regarded as an unfolding of the immediate past into the adjacent present.  Formally we can regard this process as a movement from one set of commuting variables into another commuting set of the {\em same}  variables.  What about the other non-commuting sets of variables?  In  section \ref{sec:DandF}, we saw that each infinitesimal TA was accompanied by a pair of momentum TAs, $p_\epsilon(q',q)$ and $p_\epsilon'(q',q)$.  How do  we relate these TAs to a particle moving between $q$ and $q'$?
  
  To avoid problems with the uncertainty principle, consider a small but finite volume $\Delta V$ surrounding the point $q$.  Imagine a sequence of particles emanating from a point in  $\Delta V$, each with a different momentum, so that over time we have a spray of possible momenta emerging from the volume $\Delta V$.  Similarly there is a spray of momenta over time arriving at the small volume $\Delta V'$ surrounding the point $q'$.  
  
\begin{figure}[h] 
     \centering
     \includegraphics[width=4in]{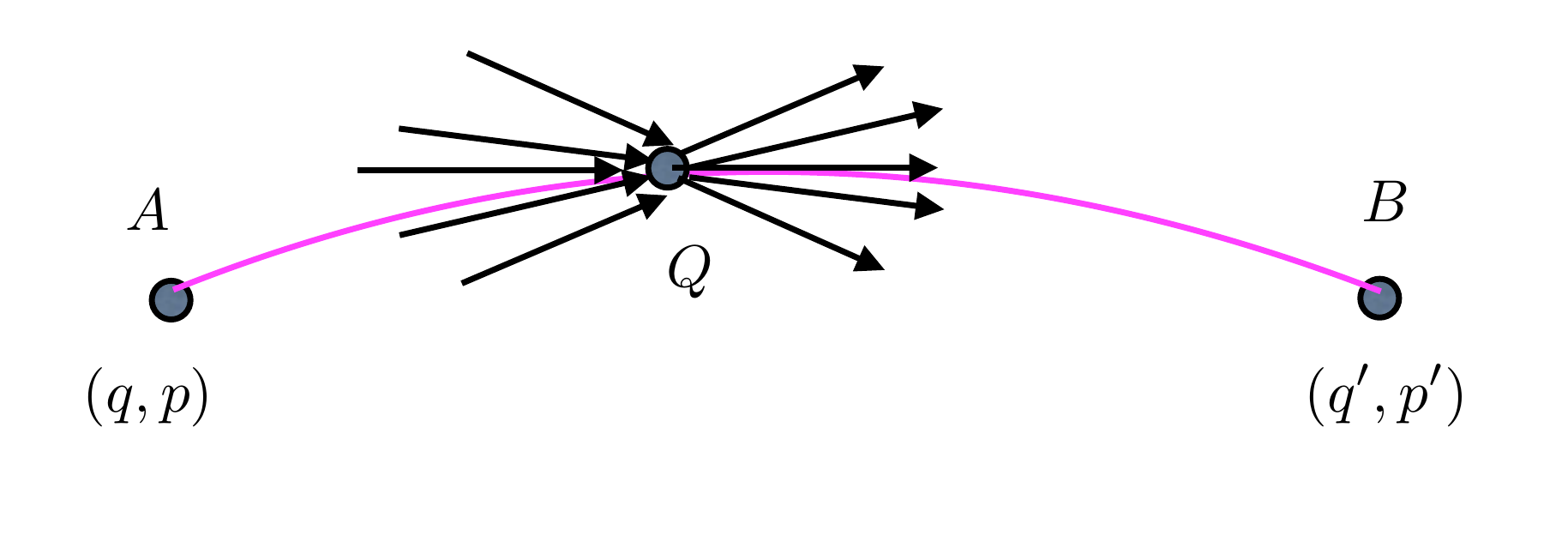} 
     \caption{Behaviour of the momenta sprays at the midpoint of $\langle q', t'| q, t\rangle_\epsilon$}
     \label{fig:spray}
  \end{figure}

  Better still, let us consider a small volume surrounding the midpoint $Q$.  At this point there is a spray of momenta arriving and a spray leaving a volume $\Delta V(Q)$ as shown in Figure \ref{fig:spray}.
  To see how the local momenta behave at the midpoint $Q$, recall that for small time differences $t'-t=\epsilon$, we have for the propagator of a free particle (\ref{eq:lift}),
  \begin{eqnarray}
  S_\epsilon(q',q)=\frac{m}{2}\frac{(q'-q)^2}{\epsilon}	\label{eq:act}
  \end{eqnarray}
  which is obtained from the classical Lagrangian.  Then we have the  momentum TA
  \begin{eqnarray}
  P_Q(q',q)=\frac{\partial S_\epsilon(q', q)}{\partial Q}=\frac{\partial S_\epsilon(Q,q)}{\partial Q}+\frac{\partial S_\epsilon(q', Q)}{\partial Q}.	\label{eq:momXa}
  \end{eqnarray}
  Using  (\ref{eq:act}), we find
  \begin{equation}
  P_Q(q',q)=m\left[\frac{(q'-Q)}{\epsilon}-\frac{(Q-q)}{\epsilon}\right]=p'_Q(q',q)+p_Q(q',q).  \label{eq:momX}
  \end{equation}
  The RHS of this equation comprises exactly the relations that Dirac~\cite{pd45} obtained in equations (\ref{eq:conjM1}) and  (\ref{eq:conjM}) above.  Not surprisingly, it is also
   exactly the momentum TA that Feynman~\cite{rpf48} obtains in his equation (48)  at the point $Q$ which lies between the two neighbouring points separated in time by $\Delta t=\epsilon$. 
   
    Notice that in the limit of $\epsilon\rightarrow 0$, equation (\ref{eq:momX}) comprises two `derivatives' at $Q$, namely
  \begin{eqnarray*}
  D_Q(\mbox{Backward})=\lim_{\epsilon\rightarrow 0}\frac{(Q-q)}{\epsilon}\hspace{1cm}  D_Q(\mbox{Forward})=\lim_{\epsilon\rightarrow 0}\frac{(q'-Q)}{\epsilon}.
  \end{eqnarray*}
  Such derivatives are associated with a general stochastic process where the `trajectory' joining the two points $q$ to $q'$ is continuous, but the derivatives are not.   This situation is known to arise in Brownian motion~\cite{nw66}.  Indeed these very derivatives were used by Nelson~\cite{en66}  in his derivation of the Schr\"{o}dinger equation from an underlying stochastic process. (See also the discussion in Bohm and Hiley~\cite{dbbh89} and  Prugove\u{c}ki~\cite{ep82} for alternative views.)

The meaning of the non-continuous derivatives here is clear;
the basic underlying quantum process connecting infinitesimally neighbouring points is an {\em intrinsically} random process, but at this stage the precise form of this stochastic process is unclear.  However the spray of possible momenta emanating from a region cannot be completely random since, as Feynman has shown, the transition amplitudes satisfy the Schr\"{o}dinger equation under certain assumptions.  Some clues as to the precise nature of this distribution have already been supplied by Takabayasi~\cite{tt54} and Moyal~\cite{jm49}, clues which we will now exploit.

We are interested in finding  the average behaviour of the momentum, $P_Q$, at the point $Q$.  This means we must determine the spray of momenta that is consistent with the wave function $\psi(Q)$ at $Q$.  But we have two contributions, one coming from the point $q$ and one leaving for the point $q'$.   Feynman's proposal~\cite{rpf48} that we can think of $\psi(Q)$ as `information coming from the past' and $\psi^*(Q)$ as `information coming from the future', will be used here as this suggests that we can write
\begin{eqnarray*}
\lim_{q\rightarrow Q}\psi(q)=\int\phi(p)e^{ipQ} dp\hspace{0.5cm}\mbox{and}\hspace{0.5cm} \lim_{Q\rightarrow q'}\psi^*(q')=\int \phi^*(p')e^{-ip'Q}dp'.
\end{eqnarray*}
The $\phi(p)$ contains information regarding the probability distribution of the incoming momentum spray, while $\phi^*(p')$ contains information about the  probability distribution of the outgoing momentum spray.  These wave functions must be such that in the limit $\epsilon\rightarrow 0$ they are consistent with the wave function $\psi(Q)$.
Thus we can define the mean momentum, $\overline {P(Q)}$, at the point $Q$ as
\begin{eqnarray}
\rho(Q)\overline {P(Q)}=\int\int P\phi^*(p')e^{-ip'Q}\phi(p)e^{ipQ}\delta(P-(p'+p)/2)dPdpdp'		\label{eq:MMPP}
\end{eqnarray}
where $\rho(Q)$ is the probability density at $Q$.   We have added the restriction $\delta(P-(p'+p)/2)$ because we are using the diffeomorphism $(p,p')\rightarrow [(p'+p)/2,(p'-p)]$.  It is immediately seen that equation (\ref{eq:MMPP}) can be put in the form   
 \begin{eqnarray}
    \rho(Q)\overline {P(Q)}=\left(\frac{1}{2i}\right)[(\partial_{q_1}-\partial_{q_2})\psi(q_1)\psi(q_2)]_{q_1=q_2=Q},	\label{eq:MMXX}
  \end{eqnarray}
  a form that appears in Moyal~\cite{jm49}.
  
  If we write the wave function in polar form, we find that  $\overline {P(Q)}$ is just the local momentum $P_B=\nabla S$ that appears in the Bohm interpretation.  Since $P_B$ is used to calculate the Bohm trajectories, there must be a close relationship between these trajectories and Feynman paths.  If we assume each evolving quantum process, which we will call a particle, actually follows a Feynman stochastic path then a Bohm trajectory can be regarded as an ensemble average of many such paths. Notice however, this gives a very different picture of the Bohm momentum  from the usual one used in Bohmian mechanics~\cite{ddst09}.  It is not the
momentum of a single `particle' passing the point $Q$, but the mean {\em momentum flow} at the point in question.  

This conclusion is supported by the treatment of the electromagnetic field  by Flack and Hiley~\cite{rfbh16} where the notion of a
 photon replaces that of the particle.  In their paper it was  shown  that the `photon' flow lines constructed in the experiments of Kocsis {\em et al.}~\cite{skbbsr11} were actually statistical averages producing momentum flow lines which corresponded to those determined by the {\em weak} value of the Poynting vector.  This agrees with standard quantum electrodynamics,  where the notion of a `photon trajectory' has no meaning. 
 
 \section{Conclusion}
 
The algebraic approach to quantum phenomena proposed by  Dirac~\cite{pd45} and the apparently very different approach of Bohm~\cite{db52}, are two aspects of a deeper structure which we have called the Dirac-Bohm picture. Just as in the Heisenberg picture, the operators become time dependent through a unitary transformation.  However in the Dirac-Bohm picture, the unitary transformation involves  the  classical action rather than the total energy.  This has the effect of producing a quantum formalism that is closer to the classical description, enabling one to see the essential differences between classical and quantum phenomena and providing a basis for deformation quantum mechanics~\cite{ahph02} as we will show elsewhere.

\section{Acknowledgments} 
 
We would like to thank Rob Flack, Bob Callaghan and Lindon Neil for many stimulating discussions. This research was supported by the Fetzer Franklin Fund of the John E. Fetzer memorial Trust.



\bibliography{myfile}{}
\bibliographystyle{plain}

\end{document}